\documentclass[11pt]{article}
\usepackage{amsmath}
\usepackage{amssymb}
\usepackage{epsfig}
\allowdisplaybreaks[4]
\begin{document}
\title{Minimisation of a one-loop charge breaking effective potential}
\author{P.M. Ferreira \\ Dublin Institute for Advanced Studies, \\
Ireland}
\date{February, 2001} 
\maketitle
\noindent
{\bf Abstract.} We compute the field derivatives of a one-loop charge breaking
effective potential and analyse their effect in its minimisation. The impact
on charge breaking bounds on the MSSM parameters is discussed. 
\vspace{-9cm}
\begin{flushright}
DIAS-STP-01-03
\end{flushright}
\vspace{8cm}

\begin{figure}[htb]
\begin{center}
\epsfig{height=8cm,file=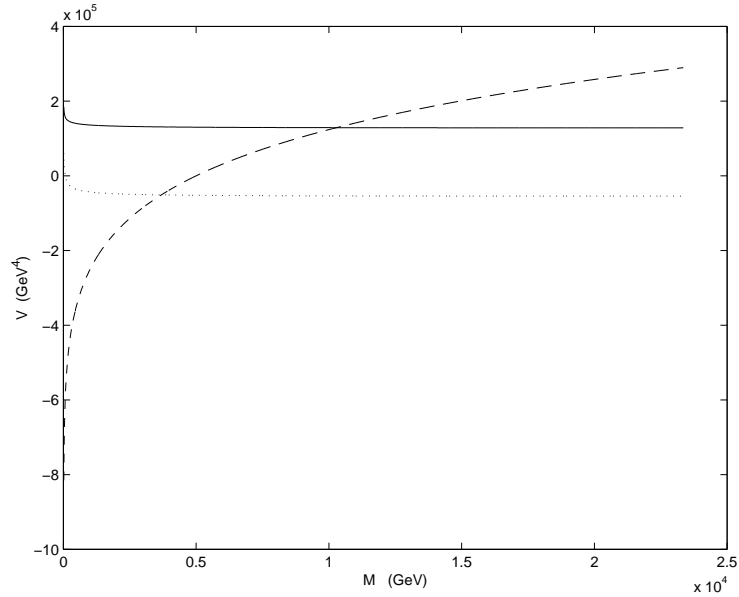}
\end{center}
\caption{Evolution of $(V_0\,+\,\Delta V_1)^{CCB}(v_i^1)$, $(V_0\,+\,\Delta 
V_1)^{CCB}(v_i^0)/100$ and $(V_0\,+\,\Delta V_1)^{MSSM}$ (solid, dashed and 
dotted lines respectively) with the renormalisation scale $M$. Notice how the 
two one-loop minimised potentials run almost parallel to one another. For this 
case, $M_G =$ 80 GeV, $m_G =$ 40 GeV, $A_G =$ 400 GeV, $\tan \beta =$ 6.5 and 
$\mu_G =$ -1.436.}
\label{fig:pot}
\end{figure}

\begin{figure}[htb]
\begin{center}
\epsfig{height=8cm,file=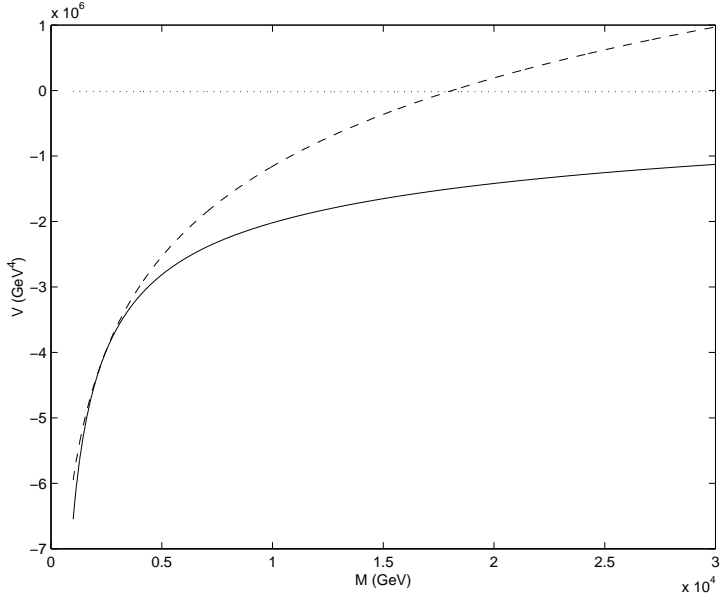}
\end{center}
\caption{Evolution of $(V_0\,+\,\Delta V_1)^{CCB}(v_i^1)$, $(V_0\,+\,\Delta
V_1)^{CCB}(v_i^0)$ and $(V_0\,+\,\Delta V_1)^{MSSM}$ (solid, dashed and
dotted lines respectively) with the renormalisation scale $M$. The strong 
variation with $M$ of the vevs displayed in fig.~\eqref{fig:vev} causes the 
similar dependence of the CCB potential here. $M_G =$ 100 GeV, $m_G =$ 160 GeV, 
$A_G =$ -600 GeV, $\tan \beta =$ 10.5 and  $\mu_G =$ -2.9823.}
\label{fig:pot2}
\end{figure}

\begin{figure}[htb]
\begin{center}
\epsfig{height=8cm,file=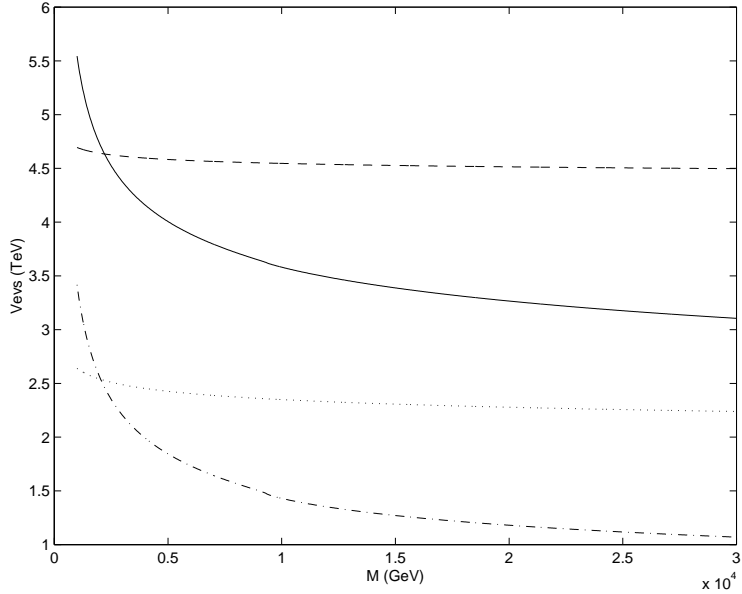}
\end{center}
\caption{Evolution of the vevs $v_1$ (from one-loop - solid line - and 
tree-level - dashed line - minimisation) and $v_2$ (from one-loop - dot-dashed
line - and tree-level - dotted line - minimisation) with the renormalisation 
scale $M$, for the same choice of parameters of fig.~\eqref{fig:pot2}. Notice 
the large difference in value between the tree-level and one-loop vevs. For very
high values of the renormalisation scale, the one-loop vevs begin to stabilise.}
\label{fig:vev}
\end{figure}

Spontaneous gauge symmetry breaking occurs in the Minimal Supersymmetric 
Standard Model when the neutral - and obviously colourless - components of the 
Higgs doublets $H_1$ and $H_2$ acquire non zero vevs. The presence in the theory
of many other scalar fields means there is no {\em a priori} reason why the
minimum of the potential should be a charge and colour preserving one. If fields
other than $H_1^0$ and $H_2^0$ have vevs and for a particular combination of 
MSSM parameters the resulting minimum is deeper than the standard one we would 
therefore be in a situation where charge and/or colour symmetries were broken. 
This simple fact gives us, in principle, a way of imposing bounds on the MSSM 
parameter space~\cite{fre}. There has been a great deal of work done in this 
area~\cite{ccb}, most of it based on the analysis of the tree-level effective 
potential along specific charge and/or colour breaking (CCB) directions. A 
fundamental point in these works is that the CCB and MSSM potentials be compared
at different renormalisation scales - this is based upon the work of Gamberini 
{\em et al}~\cite{gam}, where it was showed the vevs derived from the tree-level
MSSM potential are a reasonable approximation to those obtained from the 
one-loop potential if one chooses a renormalisation scale of the order of the 
largest particle mass in the theory. Because that typical mass is different in 
the MSSM (of the order of tens or hundreds of GeV) and the CCB (of the TeV or 
tens of TeV order) cases, comparing both potentials at the tree-level order 
should in principle be done at two different scales. The authors of 
ref.~\cite{baer} determined CCB bounds including the one-loop contributions to 
the potential from the top-stop sectors, which they argued were the most 
significant ones. 
Recently~\cite{eu} the full one-loop potential for a particular CCB direction 
was calculated and used to restrict the parameter space of the MSSM. It was 
argued that comparing the MSSM and CCB potentials at different renormalisation 
scales neglected to take into account the field-independent part of those same 
potentials, vital to ensure their renormalisation group invariance~\cite{ford}. 
The higgs and chargino contributions to the one-loop potential proved to be as 
important as the top-stop ones. This analysis was done at the typical CCB mass 
scale, so that, using the results of~\cite{gam}, the computation of the one-loop
derivatives of the CCB potential could be avoided. The results that were found 
had some renormalisation scale dependence, and it was then theorised that it 
would vanish if one performed the full one-loop minimisation of the CCB 
potential. In this letter we will undertake just that task. We rely heavily on 
the results of ref.~\cite{eu} and refer the reader to its conventions. We recall
that we only consider the Yukawa couplings of the third generation, and the 
superpotential of the model is thus given by
\begin{eqnarray}
W & = & \lambda_t H_2\,Q\,t_R \; + \; \lambda_b H_1\,Q\,b_R \; +\; \lambda_\tau
H_1\,L\,\tau_R \; +\; \mu H_2\,H_1 \;\;\; . 
\label{eq:sup}
\end{eqnarray}
Supersymmetry is broken softly in the standard manner by the inclusion in the
potential of explicit mass terms for the scalar partners and gauginos, and 
soft-breaking bilinear and trilinear terms proportional, via coefficients $A_i$
and $B$, to their counterparts in the superpotential~\eqref{eq:sup}. At a 
renormalisation scale $M$, the one-loop contributions to the potential are given
by
\begin{equation}
\Delta V_1 \, =\, \sum_\alpha \, \frac{n_\alpha}{64\pi^2}\,M_\alpha^4\, \left(\,
\log \frac{M_\alpha^2}{M^2}\, - \, \frac{3}{2}\,\right) \;\;\; ,
\label{eq:v1cor}
\end{equation}
where the $M_\alpha$ are the tree-level masses of each particle of spin
$s_\alpha$ and $n_\alpha = (-1)^{2s_\alpha}\,(2s_\alpha +1)\,C_\alpha\,
Q_\alpha$. $C_\alpha$ is the number of colour degrees of freedom  and $Q_\alpha$
is 2 for charged particles, 1 for chargeless ones. The ``real" minimum occurs
when the neutral components of $H_1$ and $H_2$ acquire vevs $v_1/\sqrt{2}$ and 
$v_2/\sqrt{2}$, the value of the tree-level potential then being
\begin{equation}
V_0 \; = \;\frac{1}{2} \left( m_1^2\,v_1^2\;+\;m_2^2\,v_2^2 \right) \;-\;B\,\mu
\,v_1\,v_2\;+\;\frac{1}{32} ({g^\prime}^2+g_2^2)\,(v_2^2-v_1^2)^2 \;\;\; ,
\end{equation}
with $m_1^2 = m_{H_1}^2+\mu^2$ and $m_2^2 = m_{H_2}^2+\mu^2$. In the CCB 
direction we will consider the scalar fields $\tau_L$ and $\tau_R$ also have
non-zero vevs, $l/\sqrt{2}$ and $\tau/\sqrt{2}$ respectively, and the vacuum 
tree-level potential is now given by
\begin{eqnarray}
V_0 & =& \frac{\lambda_\tau^2}{4} [v_1^2\,(l^2+\tau^2) + l^2\,\tau^2] \;-\;
\frac{\lambda_\tau}{\sqrt{2}}\,(A_\tau\,v_1 + \mu\,v_2)\,l\,\tau \;+ \;
\frac{1}{2} \,( m_1^2\,v_1^2 \; + \; m_2^2\,v_2^2\; + \; m_L^2\,l^2 \; +
\nonumber \\
 & & m_\tau^2\,\tau^2 )\;-\;B\,\mu\,v_1\, v_2 \;+\; \frac{g^{\prime 2}}{32}\,
(v_2^2-v_1^2
-l^2+2\,\tau^2)^2 \; +\; \frac{g_2^2}{32}\,(v_2^2-v_1^2+l^2)^2\;\;\; .   
\label{eq:vc}
\end{eqnarray}
The derivatives of this tree-level potential with respect to each of the vevs
are very simple, but the same cannot be said for the one-loop derivatives, their
total contribution given by
\begin{equation}
\sum_\alpha \, \frac{n_\alpha}{32\pi^2}\,M_\alpha^2\,\frac{\partial M_\alpha^2}{
\partial v_i}\, \left(\, \log \frac{M_\alpha^2}{M^2}\, - \, 1\,\right) \;\;\; .
\label{eq:cont}
\end{equation}
Some of the squared masses' derivatives are trivial to calculate: that is the 
case of the top and bottom quarks, and the scalar partners of the second and 
first generation up and down quarks and electron and neutrinos (expressions (15)
to (18) of ref.~\cite{eu}) - we present these results in the appendix. For the 
remaining sparticles the calculation is made more difficult by the masses being 
given by the eigenvalues of square matrices, sometimes as large as $6 \times 6$ 
- many of these matrices are much more complex than their MSSM counterparts due 
to the existence of charged vevs causing mixing of charged and neutral fields. 
For example, the ``higgs scalars" of the CCB potential are in fact the result of
the mixing between the neutral components of $H_1$, $H_2$ and the fields 
$\tau_L$ and $\tau_R$, their squared masses thus given by the eigenvalues of a 
$4\times 4$ matrix. It is nevertheless possible to find analytical expressions 
for $\partial M_\alpha^2/\partial v_i$, once the $M_\alpha$ themselves have been
determined (which is easy to do numerically, where an analytical determination 
proves impossible) - this is accomplished by noticing that the particle masses 
are always given by the roots of an $n^{th}$-order polynomial (in our case, 
$n=2$, 3, 4 and 6),
\begin{equation}
F(\lambda\;,\; v_i) \; = \; \lambda^n \; +\; A\,\lambda^{n-1} \;+\; B\,
\lambda^{n-2} \;+\; \ldots \; = \; 0 \;\;\; ,
\label{eq:F}
\end{equation}
with coefficients $\{A, B, \ldots \}$ generally depending on the vevs
$v_i=\{v_1,v_2,l,\tau\}$. This equation implicitly defines 
the squared masses $\lambda$ in function of the $v_i$, so we have
\begin{equation}
\frac{\partial M_\alpha^2}{\partial v_i} \; = \; -\; \left. \frac{{\displaystyle
\frac{\partial F}{ \partial v_i}}}{{\displaystyle \frac{\partial F}{\partial 
\lambda}}} \; = \; -\; 
\frac{{\displaystyle \frac{\partial A}{\partial v_i}}\,\lambda^{n-1} \;+\; 
{\displaystyle \frac{\partial B}{\partial v_i}}\,\lambda^{n-2} \;+\; \ldots}{n\,
\lambda^{n-1} \;+\; (n-1)\,A\, \lambda^{n-2} \;+\; (n-2)\,B\, \lambda^{n-3} \;+
\; \ldots}\;\; \right|_{\lambda = M_\alpha^2} \;\;\; .
\label{eq:imp}
\end{equation}
If $n > 2$ the final expressions depend on the numerical solving of 
eq.~\eqref{eq:F}. For $n=2$ it is possible to write down fully analytical 
expressions of the derivatives of the squared masses, but from a practical point
of view it is better to use the recipe of eq.~\eqref{eq:imp}. In the following 
we show how to calculate the coefficients $\{A,B,C,\ldots\}$ (and their 
derivatives) of eq.~\eqref{eq:F} for the several sparticles in terms of the 
elements $\{a,b,c,\ldots\}$ of their mass matrices. The derivatives of $\{a,b,c,
\ldots\}$ are listed in the appendix. So, for a symmetric $2\times 2$ mass 
matrix with diagonal elements $a$ and $c$ and off-diagonal element $b$, 
eq.~\eqref{eq:imp} reduces to
\begin{equation}
\frac{\partial M_\alpha^2}{\partial v_i} \; = \; \frac{a\,{\displaystyle
\frac{\partial c}{\partial v_i}} \;+\;c\,{\displaystyle \frac{\partial (ac)}{
\partial v_i}} \;-\;2\,b\,{\displaystyle \frac{\partial b}{\partial v_i}} \;-\;
{\displaystyle \frac{\partial (a\,+\,c)}{\partial v_i}}}{a\;+\;c\;-\;2\,\lambda}
\;\;\; .
\end{equation}
This is the case of the stop, sbottom and neutral gauge boson masses, the 
coefficients $\{a,b,c\}$ are given in eqs.~(12),~(14) and~(21) of 
ref.~\cite{eu}, 
and their derivatives a simple calculation. The squared masses of the charginos 
are also determined by a quadratic equation~\footnote{One of the eigenvalues of 
the $5 \times 5$ chargino mass matrix is zero, corresponding to the $\tau$ 
neutrino. This leaves a quartic equation in the masses, that reduces to a 
quadratic one in the squared masses.}, namely (from eq.~(22) of ref.~\cite{eu}),
$\lambda^2\;-\;A_{\chi^\pm}\,\lambda\;+\;B_{\chi^\pm}\;=\;0$, with
\begin{eqnarray}
A_{\chi^\pm} & = & M_2\;+\;\mu^2\;+\;\frac{1}{2}\,\left[ g_2^2\, (v_1^2+v_2^2+
l^2) \;+\; \lambda_\tau^2\,\tau^2\right] \nonumber \\
B_{\chi^\pm} & = & \left(\frac{1}{2}\,g_2^2\,v_1\,v_2\;-\;\mu\,M_2\right)^2\;+\;
\frac{1}{2}\,g_2^2\,l^2\,\left(\frac{1}{2}\,g_2^2\,v_1^2\;+\;\mu^2\right) \;+\;
\frac{\lambda_\tau^2}{2}\,\left(\frac{1}{2}\,g_2^2\,v_2^2\;+\;M_2^2\right)\,
\tau^2 \; -\nonumber \\
 & & \frac{\lambda_\tau}{\sqrt{2}}\,g_2^2\,(\mu\,v_2\;+\;M_2\,v_1)\,l\,
\tau \;\;\; .
\label{eq:char}
\end{eqnarray}
All that remains is to calculate the derivatives of $A_{\chi^\pm}$ and 
$B_{\chi^\pm}$ and substitute their values in eq.~\eqref{eq:imp}. The charged 
Higgses are a mix between the charged components of $H_1$ and $H_2$ and the tau 
sneutrino, with a $3 \times 3$ mass matrix with elements $\{a_\pm,b_\pm, \ldots,
f_\pm\}$, as shown in eq.~(24) of ref.~\cite{eu}. The squared masses end up 
being determined by a cubic equation, 
\begin{equation}
\lambda^3\;-\;A_\pm\,\lambda^2\;-\;B_\pm\,\lambda\;-\;C_\pm\;=\;0\;\;\; ,
\end{equation}
with
\begin{eqnarray}
A_\pm & = & a_\pm \;+\;c_\pm\;+\;f_\pm \nonumber \\ 
B_\pm & = & b_\pm^2\;+\;d_\pm^2\;+\;e_\pm^2\;-\;a_\pm\,(c_\pm+f_\pm)\;-\;c_\pm\,
f_\pm \;\;\; .
\end{eqnarray}
$C_\pm$ is, of course, the determinant of the mass matrix, but we end up not
needing to calculate it, only its derivative. Adopting the convention $X_{,v}$
to indicate $\partial X/\partial v$, with $v$ any of the vevs, we obtain
\begin{eqnarray}
A_{\pm,v} & = & a_{\pm,v} \;+\;c_{\pm,v}\;+\;f_{\pm,v} \nonumber \\
B_{\pm,v} & = & 2\,(b_\pm\,b_{\pm,v}\;+\;d_\pm\,d_{\pm,v}\;+\;e_\pm\,_\pm\,
e_{\pm,v})\;-\;a_\pm\,(c_{\pm,v}\;+\;f_{\pm,v})\;-\;c_\pm\,(a_{\pm,v}\;+\; 
f_{\pm,v})\;-  \nonumber \\
 & & f_\pm\,(a_{\pm,v}\;+\;c_{\pm,v}) \nonumber \\
C_{\pm,v} & = & 2\,(b_{\pm,v}\,d_\pm\,e_\pm\;+\;d_{\pm,v}\,b_\pm\,e_\pm\;+\;
e_{\pm,v}\,b_\pm\,d_\pm)\;+\;a_{\pm,v}\,c_\pm\,f_\pm\;+\;c_{\pm,v}\,a_\pm\,f_\pm
\;+ \nonumber \\
 & & f_{\pm,v}\,a_\pm\,c_\pm\;-\;d_\pm\,(2\,d_{\pm,v}\,c_\pm\;+\;c_{\pm,v}\,
d_\pm)\;-\;b_\pm\,(2\,b_{\pm,v}\,f_\pm\;+\;f_{\pm,v}\,b_\pm)\;-\nonumber \\
 & & e_\pm\,(2\, e_{\pm,v}\,a_\pm\;+\;a_{\pm,v}\,e_\pm) \;\;\; .
\end{eqnarray}
The squared masses of both the pseudoscalars and Higgs scalars are given by $4
\times 4$ matrices (eqs.~(26) and~(28) of ref.~\cite{eu}) with one set of 
coefficients $\{ a, b, \ldots j\}$ for each case. The resulting fourth-order
eigenvalue equation,
\begin{equation}
\lambda^4\;-\;A\,\lambda^3\;+\;B\,\lambda^2\;+\;C\,\lambda\;+\;D\;=\;0\;\;\; ,
\end{equation}
has coefficients
\begin{eqnarray}
A & = & a\;+\;c\;+\;h\;+\;j \nonumber \\
B & = & a\,(c\;+\;h\;+\;j)\;+\;c\,(h\;+\;j)\;+\;h\,j\;-\;b^2\;-\;d^2\;-\;e^2-\;
f^2\;- \;g^2-\;i^2 \nonumber \\
C & = & a\,(g^2\;+\;f^2\;+\;i^2\;-\;c\,j\;-\;c\,h\;-\;h\,j)\;+\;c\,(d^2\;+\;e^2
\;+\;i^2\;-\;h\,j)\;+ \nonumber \\
 & & h\,(b^2\;+\;e^2\;+\;g^2)\;+\;j\,(b^2\;+\;d^2\;+\;f^2)\;- \nonumber \\
 & & 2\,(f\,g\,i\;+\; b\,f\,d\;+\;d\,e\,i\;+\;b\,g\,e) \;\;\; .
\end{eqnarray}
Once again, for the calculation of the derivatives of eq.~\eqref{eq:imp}, we 
don't need the explicit value of $D$, the determinant of the mass matrix. With 
the same convention as before, we have
\begin{eqnarray}
A_{,v} & = & a_{,v}\;+\;c_{,v}\;+\;h_{,v}\;+\;j_{,v} \nonumber \\
B_{,v} & = & a\,(c_{,v}\;+\;h_{,v}\;+\;j_{,v})\;+\;c\,(a_{,v}\;+\;h_{,v}\;+\;
j_{,v})\;+\;h\,(a_{,v}\;+\;c_{,v}\;+\;j_{,v})\;+  \nonumber \\
 & & j\,(a_{,v}\;+\;c_{,v}\;+\; h_{,v})\;-\;2\,(b\,b_{,v}\;+\;d\,d_{,v}\;+\;e\,
e_{,v}\;+\;f\,f_{,v}\;+\;g\,g_{,v}\;+\;i\,i_{,v}) \nonumber \\
C_{,v} & = & a_{,v}\,(g^2\,+\,f^2\,+\,i^2\,-\,c\,j\,-\,c\,h\,-\,h\,j)\,+\,c_{,v}
\,(d^2\,+\,e^2\,+\,i^2\,-\,a\,j\,-\,a\,h\,-\,h\,j)\,+ \nonumber \\
 & & h_{,v}\,(b^2\,+\,e^2\,+\,g^2-\,a\,j\,-\,a\,c\,-\,c\,j)\,+\,j_{,v}\,(b^2\,+
\,d^2\,+\,f^2-\,a\,h\,-\,a\,c\, -\,c\,h)\,+ \nonumber \\
 & & 2\,b_{,v}\,(b\,h\,+\,b\,f\,-\,f\,d\,-\,g\,e)\,+\,2\,d_{,v}\,(d\,c\,+\,d\,j
\,-\, b\,f\,-\,e\,i)\,+ \nonumber \\
 & & 2\,e_{,v}\,(e\,c\,+\,e\,h\,-\,d\,i\,-\,b\,g)\,+\,2\,f_{,v}\,(f\,a\,+\,f\,j
\,-\,g\,i\,-\,b\,d)\,+ \nonumber \\
 & & 2\,g_{,v}\,(g\,a\,+\,g\,h\,-\,f\,i\,-\,e\,b)\,+\,2\,i_{,v}\,(i\,a\,+\,i\,c
\,-\,g\,f\,-\,e\,d) \nonumber \\
D_{,v} & = & a_{,v}\,(c h j+2\,f g i-j f^2-c i^2-h g^2)\,+\, 
c_{,v}\,(a h j + 2\,d e i-j d^2-a i^2-h e^2) \,+
\nonumber \\
 & & h_{,v}\,(a c j+2\,beg-jb^2-ag^2-ce^2)\,+\,j_{,v}\,(ach+2\,bfd-hb^2-af^2-
cd^2)\,+ \nonumber \\
 & & 2\,b_{,v}\,(b\,i^2\,-\,b\,h\,j\,+\,f\,d\,j\,+\,g\,e\,h\,-\,f\,e\,i\,-\,g\,d
\,i) \,+ \nonumber \\
 & & 2\,d_{,v}\,(d\,g^2\,-\,c\,d\,j\,+\,f\,b\,j\,+\,c\,e\,i\,-\,b\,g\,i\,-\,e\,g
\,f)\,+ \nonumber \\
 & & 2\,e_{,v}\,(e\,f^2\,-\,c\,h\,e\,+\,g\,b\,h\,+\,c\,d\,i\,-\,b\,f\,i\,-\,
g\,d\,f)\,+ \nonumber \\
 & & 2\,f_{,v}\,(f\,e^2\,-\,a\,f\,j\,+\,d\,b\,j\,+\,a\,g\,i\,-\,b\,e\,i
\,-\,g\,d\,e)\,+ \nonumber \\
 & & 2\,g_{,v}\,(g\,d^2\,-\,a\,h\,g\,+\,e\,b\,h\,+\,a\,f\,i\,-\,b\,
d\,i\,-\,f\,d\,e)\,+ \nonumber \\
 & & 2\,i_{,v}\,(i\,b^2\,-\,a\,c\,i\,+\,a\,f\,g\,+\,d\,c\,e\,- \,b\,f\,e\,-\,b\,
d\,g)
\end{eqnarray}
For the neutralinos, the sixth-order equation for the masses is made reasonably
simple by the mass matrix (eq.~(23) of~\cite{eu}) having several zeroes. We
thus have
\begin{equation}
\lambda^6\;-\;A_{\chi^0}\,\lambda^5\;-\;B_{\chi^0}\,\lambda^4\;+\;C_{\chi^0}\,
\lambda^3\;+\;D_{\chi^0}\,\lambda^2\;+\;E_{\chi^0}\,\lambda\;+\;F_{\chi^0}\,
\lambda\;=\;0\;\;\; ,
\end{equation}
with
\begin{eqnarray}
A_{\chi^0} & = & M_1\,+\,M_2 \nonumber \\
B_{\chi^0} & = & \frac{\lambda_\tau^2}{2}\,(v_1^2\,+\,l^2\,+\,\tau^2)\;+\;\frac{
1}{4}\,({g^\prime}^2\,+\,g_2^2)\,(v_1^2\,+\,v_2^2\,+\,l^2)\;+\;{g^\prime}^2\,
\tau^2\;+\;\mu^2\;-\;M_1\,M_2 \nonumber \\
C_{\chi^0} & = & -\,\frac{\lambda_\tau^2}{2}\,(M_1\,+M_2)\,(v_1^2\,+\,l^2\,+\,
\tau^2)\;-\;\frac{1}{4}\,({g^\prime}^2\,M_2\,+\,g_2^2\,M_1)\,(v_1^2\,+\,v_2^2
\,+\,l^2) \;-  \nonumber \\
 & & {g^\prime}^2\,M_2\,\tau^2\;+\;\frac{1}{2}\,({g^\prime}^2\,+\,g_2^2)\,\mu
\,v_1\,v_2\;-\;\frac{\lambda_\tau}{2\sqrt{2}}\,(3\,{g^\prime}^2\,-\,g_2^2\,-\,2
\,\lambda_\tau^2)\,v_1\,l\,\tau\;-\;\mu^2\,(M_1\,+\,M_2) \nonumber \\
D_{\chi^0} & = &\frac{\lambda_\tau^3}{\sqrt{2}}\,(M_1\,+\,M_2)\,v_1\,l\,\tau\;+
\;\lambda_\tau^2\,\left\{\frac{}{} \right. \frac{{g^\prime}^2}{2}\,\tau^2\,
(v_1^2\,+\,l^2\,+\,\tau^2)\;+\;\frac{1}{8}\,({g^\prime}^2\,+\,g_2^2)\,\left[
v_1^4\,+\,l^4\,+ \right. \nonumber \\
 & & \left. l^2\,(v_2^2\,-\,2\,v_1^2)\,+\,v_2^2\,(v_1^2\,+\,\tau^2)\right]
\;-\;\frac{1}{2}\,M_1\,M_2\,(v_1^2\,+\,l^2\,+\,\tau^2)\,+\frac{\mu^2}{2}\,v_1^2
\left. \frac{}{}\right\} \;+ \nonumber \\
 & & \frac{\lambda_\tau}{2\sqrt{2}}\,l\,\tau\,\left[(g_2^2\,M_1\,-\,3
\,{g^\prime}^2\,M_2)\,v_1\;+\;({g^\prime}^2\,-\,g_2^2)\,\mu\,v_2\right]\;+
\nonumber \\
 & &\frac{1}{4}\,{g^\prime}^2\,g_2^2\,(v_1^2\,+\,v_2^2\,+\,l^2)\,\tau^2\;+\;
\frac{1}{2}\,({g^\prime}^2\,M_2\,+\,g_2^2\,M_1)\,\mu\,v_1\,v_2\;+ \nonumber \\
 & & \left[\frac{1}{4}\,({g^\prime}^2\,+\,g_2^2)\,l^2\;+\;{g^\prime}^2\,\tau^2
\;-\;\, M_1\,M_2\right]\,\mu^2 \nonumber \\
E_{\chi^0} & = & \frac{\lambda_\tau^3}{\sqrt{2}}\,\left[\frac{1}{4}\,(
{g^\prime}^2\,+\, g_2^2)\,v_2^2\;-\;M_1\,M_2\right]\,v_1\,l\,\tau\;+\;
\lambda_\tau^2\,\left\{\frac{}{} \right. -\, \frac{1}{8}\, ({g^\prime}^2\,M_2\,+
\,g_2^2\,M_1)\, \left[v_1^4\,+\,l^4\,+ \right. \nonumber \\
 & & \left. l^2\,(v_2^2\,-\,2\,v_1^2)\,+\,v_2^2\,(v_1^2\,+\,\tau^2)\right]\;-\;
\frac{1}{4}\,({g^\prime}^2\,+\,g_2^2)\,\mu\,v_1\,v_2\,(l^2\,-\,v_1^2)\;- 
\nonumber \\
 & & \frac{\mu^2}{2}\,(M_1\,+\,M_2)\,v_1^2\;-\;\frac{{g^\prime}^2}{2}\,
\tau^2\, \left[M_2\,(v_1^2\,+\,l^2\,+\,\tau^2)\;-\;\mu\,v_1\,v_2\right]\left.
\frac{}{} \right\}\; +\nonumber \\
 & & \frac{\lambda_\tau}{2\sqrt{2}}\,l\,\tau\,\left[{g^\prime}^2\,(g_2^2\,\tau^2
\;-\;2\,\mu^2)\,v_1\;+\;(g_2^2\,M_1\,-\,{g^\prime}^2\,M_2)\,\mu\,v_2\right]\;-
\nonumber \\
 & & \frac{1}{4}\,({g^\prime}^2\,M_2\,+\,g_2^2\,M_1)\,\mu^2\,l^2\;+\;
\frac{{g^\prime}^2}{2}\,(g_2^2\,v_1\,v_2\;-\;2\,\mu\,M_2)\,\mu\,\tau^2 \nonumber
 \\
F_{\chi^0} & = & -\,\frac{\lambda_\tau^3}{4\sqrt{2}}\,({g^\prime}^2\,M_2\,+\,
g_2^2\,M_1) \,v_1\,l\,\tau\,v_2^2\;+\;\frac{\lambda_\tau^2}{4}\,\left[(
{g^\prime}^2\,M_2\,+ \,g_2^2\,M_1)\,\mu\,(l^2\,-\,v_1^2)\,v_1\,v_2\;+ \frac{}{} 
\right. \nonumber \\
 & & \left. 2\,\mu^2\,M_1\,M_2\,v_1^2\;-\;\frac{{g^\prime}^2}{2}\,v_2\,\tau^2\,
(4\,\mu\,M_2\,v_1\;+\;g_2^2\,\tau^2\,v_2)\right]\;+ \nonumber \\
 & & \frac{\lambda_\tau}{2\sqrt{2}}\,{g^\prime}^2\,(2\,\mu\,M_2\,v_1\;+\;g_2^2\,
\tau^2\,v_2)\,\mu\,l\,\tau\;-\;\frac{\mu^2}{4}\,{g^\prime}^2
\,g_2^2\,l^2\,\tau^2 \;\;\; .
\label{eq:neut}
\end{eqnarray}
With such complex formulae, a check of the results is quite useful - because
supersymmetry is softly broken, $Str\,M^2$ is a field-independent quantity, so
we should have $Str\,\partial M^2/\partial v_i\,=\,0$. In this manner we can 
check the mass matrices themselves~\footnote{Verifying that $Str\,M^2$ is 
field-independent provides a check only on the diagonal elements of the mass 
matrices (except for the neutralinos). But because we are using 
eq.~\eqref{eq:imp} to calculate the mass derivatives, this second check involves
all the coefficients of the mass matrices.} and the consistency of our sign 
conventions. With the derivatives~\eqref{eq:cont} computed, we can perform the 
one-loop minimisation of the CCB potential. We apply our formulae to the same 
simple model of ref.~\cite{eu}: one with universality of the soft parameters at 
the gauge unification scale and input parameters in the ranges $20 \leq M_G\leq 
100$ GeV, $10 \leq m_G \leq 160$ GeV, $-600 \leq A_G \leq 600$ GeV and $2.5 \leq
\tan \beta \leq 10.5$, with both signs of the $\mu$ parameter considered. In the
work of ref.~\cite{eu}, we had performed the tree-level minimisation of the CCB
potential at a renormalisation scale $M = 0.6 \,g_2 |A_\tau|/\lambda_\tau$ - 
which was shown to be of the order of the highest of the CCB masses - and, out
of an initial parameter space of about 3200 ``points", CCB extrema had been 
found for roughly $40 \%$ of the cases. However, by repeating the process with 
the full one-loop derivatives of the CCB potential, we encounter drastically 
different results - only in almost 200 ``points" does CCB {\em seem} to occur. 
These ``points" are uniformly distributed according to the input parameters of 
$M_G$ and $m_G$, do not occur for values of $\mu_G$ close to zero (like in 
ref.~\cite{eu}) and occur mostly for $\tan \beta >$ 6 and $150<|A_G|<500$ GeV.
In figure~\eqref{fig:pot} we see the reason for the discrepancy between these 
results and those of ref.~\cite{eu}: there we have plotted the evolution 
with the renormalisation scale $M$ of the value of the one-loop MSSM potential 
($(V_0+\Delta V_1)^{MSSM}$, with one-loop vevs) and the one-loop CCB potentials 
calculated with the tree-level CCB vevs ($(V_0+ \Delta V_1)^{CCB}(v_i^0)$ - for 
convenience, we divide it by a factor of 100) and the one-loop derived vevs 
($(V_0+\Delta V_1)^{CCB}(v_i^1)$). To interpret this figure, we need to remember
that $V_0+\Delta V_1$ is {\em not} a one-loop renormalisation scale independent 
quantity~\cite{ford}, rather, in terms of the parameters $\lambda_i$ and fields 
$\phi_j$, the RGE invariant effective potential is given by
\begin{equation}
V(M,\lambda_i,\phi_j) \; =\; \Omega(M,\lambda_i) \;+\; V_0(\lambda_i,\phi_j) \;
+\; \Delta V_1 (M,\lambda_i,\phi_j)\; +\; O(\hbar^2) \;\;\; .
\label{eq:om}
\end{equation}
The only difference between the CCB and MSSM potentials is the different set of
values for some of the fields $\phi_j$, which means the field-independent 
function $\Omega$ is the same in both cases~\footnote{This was the argument used
in ref.~\cite{eu} to argue that both potentials should be compared at the same 
renormalisation scale.}. Therefore, given that $V$ is renormalisation scale
invariant, we must have $d(V_0+\Delta V_1)^{CCB}/d M \;=\; d(V_0+\Delta 
V_1)^{MSSM}/d M$ - this is certainly the case for the two one-loop minimised 
potentials of fig.~\eqref{fig:pot} (a plot of their renormalisation scale 
derivatives would show them to be almost identical for $M \gtrsim 1$ TeV), as 
they run parallel to one another, but the one-loop potential 
calculated with the tree-level vevs is clearly different. It has a very strong 
$M$ dependence, and the inequality $(V_0+\Delta V_1)^{CCB}(v_i^0)\,<\,(V_0+
\Delta V_1)^{MSSM}$ is verified only for $M \lesssim 4$ TeV. Notice how, judging
by the value of the one-loop minimised CCB potential, this ``point" is not a CCB
minimum. Unfortunately, we find that for those points that are identified as 
one-loop CCB minima the potential does not have the correct renormalisation 
scale dependence seen in fig.~\eqref{fig:pot}, as may be seen in 
fig.~\eqref{fig:pot2} - there, for a different choice of parameters, we obtain 
CCB potentials that, whether computed with tree-level or one-loop vevs, are
strongly dependent on $M$. Although the one-loop vevs do seem to somewhat
stabilize $(V_0+\Delta V_1)^{CCB}$, one can expect its value will become greater
than the MSSM potential for a higher renormalisation scale. Again, the finding
of a CCB minimum becomes $M$-dependent. The reason for this seems to be the 
values of the vevs found - in the case of fig.~\eqref{fig:pot} the one-loop vevs
are smaller than 1 TeV, and very stable against variations in $M$. But, for the 
``points" where we find $(V_0+\Delta V_1)^{CCB}(v_i^1)\,<\,(V_0+\Delta 
V_1)^{MSSM}$, the values of the vevs are much bigger than in the previous case 
and, as may be seen in fig.~\eqref{fig:vev}, change immensely with the 
renormalisation scale. In the same plot we also see the evolution of the 
tree-level vevs - remarkably they are rather stable with $M$, but the potential 
thereof resulting is still strongly $M$ dependent. We must compare this figure 
to fig.~(2.b) of the work of Gamberini {\em et al}~\cite{gam}: whereas there, 
for a range of $M$ of the order of the largest mass present in $\Delta V_1$, the
tree-level and one-loop vevs coincide, in our CCB potential they simply touch in
one particular point. We must add that for the seemingly perfect case of 
fig.~\eqref{fig:pot} the vevs do not even touch: $v_1^0$ had a fairly stable 
value of about 4.6 TeV for the whole range of $M$, and $v_1^1$, also stable, was
equal to $\sim 0.8$ TeV. 

In conclusion, we have shown that the one-loop contributions to the minimisation
of a CCB potential have a large effect in both the values of the vevs and the
potential itself. We found that the one-loop minimisation stabilizes both vevs
and the potential against changes in the renormalisation scale, but only if the
values of the vevs found are small (``small" in this case being inferior to 
about 1 TeV). In those cases, however, no CCB minima are found. We did find a
small number of CCB minima, but the corresponding vevs had large values and the 
one-loop potential proved to be strongly $M$ dependent. We believe the reason 
for this difference in behaviour is a breakdown in perturbation theory - for
higher values of the fields, the one-loop contributions become too large and 
two-loop terms become necessary to achieve renormalisation scale invariance. As
a result, the CCB minima found cannot be trusted - the corresponding value of 
the potential may well be smaller than the MSSM one, but perturbation theory is
no longer valid and it is altogether possible the two-loop contributions would
reverse that result. Also very important is the fact the tree-level derived vevs
do not coincide with the one-loop ones for a range of renormalisation scale in 
the way described in ref.~\cite{gam}. The conclusion is that taking a 
renormalisation scale of the order of the largest mass present in $\Delta V_1$ 
does not correctly reproduce the effect of the one-loop contributions to the 
potential, at least for this particular CCB potential and these values for the 
SUSY parameters - as the usual CCB analysis rely on this assumption, our 
findings cast doubt over their validity. Overall, the importance of 
performing a one-loop minimisation whilst studying CCB bounds cannot be 
underestimated, even if the results are not to our liking. To avoid the 
perturbative breakdown we encountered here we should study a CCB direction with 
lower typical masses, which suggests those directions associated with the top 
Yukawa coupling. We expect to approach this subject in the future. 
\appendix
\section{Appendix}
We now list the non-zero derivatives with respect to the vevs $\{v_1,v_2,l,\tau
\}$ of squared masses and elements of mass matrices. For the top and bottom 
quarks, the scalar quarks and leptons of the first and second generations and 
the charged gauge bosons, we have
\begin{align}
M_{t,v_2}^2 & = \lambda_t^2\,v_2 & 
M_{b,v_1}^2 & = \lambda_b^2\,v_1 &
M^2_{\tilde{u}_1,v_1} & = -\frac{1}{12}({g^\prime}^2-3\,g_2^2)\,v_1 \nonumber \\
M^2_{\tilde{u}_1,v_2} & = \frac{1}{12}({g^\prime}^2-3\,g_2^2)\,v_2 & 
M^2_{\tilde{u}_1,l} & = -\frac{1}{12}({g^\prime}^2+3\,g_2^2)
\,l  &
M^2_{\tilde{u}_1,\tau} & = \frac{1}{6}\,{g^\prime}^2\,\tau \nonumber \\ 
M^2_{\tilde{u}_2,v_1} & = \frac{1}{3}\,{g^\prime}^2\,v_1 & 
M^2_{\tilde{u}_2,v_2} & = -\frac{1}{3}\,{g^\prime}^2\,v_2 &
M^2_{\tilde{u}_2,l} & = \frac{1}{3}\,{g^\prime}^2\,l \nonumber \\
M^2_{\tilde{u}_2,\tau} & = -\frac{2}{3}\,{g^\prime}^2\,\tau &
M^2_{\tilde{d}_1,v_1} & = -\frac{1}{12}({g^\prime}^2+3\,g_2^2)\,v_1 &
M^2_{\tilde{d}_1,v_2} & = \frac{1}{12}({g^\prime}^2+3\,g_2^2)
\,v_2 \nonumber \\
M^2_{\tilde{d}_1,l} & =  \frac{1}{12}(-{g^\prime}^2+3\,g_2^2)\, l & 
M^2_{\tilde{d}_1,\tau} & = \frac{1}{6}\,{g^\prime}^2\,\tau & 
M^2_{\tilde{d}_2,v_1} & = -\frac{1}{6}\,{g^\prime}^2\,v_1 \nonumber \\ 
M^2_{\tilde{d}_2,v_2} & = \frac{1}{6}\,{g^\prime}^2\,v_2 & 
M^2_{\tilde{d}_2,l} & = -\frac{1}{6}\,{g^\prime}^2\,l & 
M^2_{\tilde{d}_2,\tau} & = \frac{1}{3}\,{g^\prime}^2\,\tau \nonumber \\
M^2_{\tilde{e}_1,v_1} & = -\frac{1}{4}(g_2^2-{g^\prime}^2)\,v_1 & 
M^2_{\tilde{e}_1,v_2} & = \frac{1}{4}(g_2^2-{g^\prime}^2)\,v_2 
& M^2_{\tilde{e}_1,l} & = \frac{1}{4}(g_2^2+{g^\prime}^2)\,l \nonumber \\
M^2_{\tilde{e}_1,\tau} & = -\frac{1}{2}{g^\prime}^2\,\tau & 
M^2_{\tilde{e}_2,v_1} & = -\frac{1}{2}{g^\prime}^2\,v_1 &
M^2_{\tilde{e}_2,v_2} & = \frac{1}{2}{g^\prime}^2\,v_2 \nonumber \\
M^2_{\tilde{e}_2,l} & = -\frac{1}{2}{g^\prime}^2\,l &
M^2_{\tilde{e}_2,\tau} & = {g^\prime}^2\,\tau & 
M^2_{\tilde{\nu}_e,v_1} & = \frac{1}{4}({g^\prime}^2+g_2^2)\,v_1 \nonumber \\
M^2_{\tilde{\nu}_e,v_2} & = -\frac{1}{4}({g^\prime}^2+g_2^2)\,v_2 & 
M^2_{\tilde{\nu}_e,l} & = \frac{1}{4}({g^\prime}^2-g_2^2)\,l & 
M^2_{\tilde{\nu}_e,\tau} & = -\frac{1}{2}{g^\prime}^2\,\tau \nonumber \\ 
M^2_{W,v_1} & = \frac{1}{2}\,g_2^2\,v_1 &
M^2_{W,v_2} & = \frac{1}{2}\,g_2^2\,v_2 &
M^2_{W,l} & = \frac{1}{2}\, g_2^2\,l 
\end{align}
For the stop, sbottom and neutral gauge bosons, whose squared masses are given
by symmetric $2 \times 2$ matrices with diagonal elements $a$ and $c$ and
off-diagonal element $b$, we have
\begin{align}
a_{\tilde{t},v_1} & = -\frac{1}{12}({g^\prime}^2-3\,g_2^2)\, v_1 & 
a_{\tilde{t},v_2} & = \frac{1}{12}(12\,\lambda_t^2+{g^\prime}^2-3\,g_2^2)\,v_2 &
a_{\tilde{t},l} & = -\frac{1}{12}({g^\prime}^2+3\,g_2^2)\,l 
\nonumber \\
a_{\tilde{t},\tau} & = \frac{1}{6}\,{g^\prime}^2\,\tau & 
b_{\tilde{t},v_1} & = \frac{\lambda_t}{\sqrt{2}}\,\mu & 
b_{\tilde{t},v_2} & = -\frac{\lambda_t}{\sqrt{2}}\,A_t \nonumber \\
c_{\tilde{t},v_1} & = \frac{1}{3}\,{g^\prime}^2\,v_1 &
c_{\tilde{t},v_2} & = \frac{1}{3}(3\,\lambda_t^2-\,{g^\prime}^2)\,v_2
 & 
c_{\tilde{t},l} & = \frac{1}{3}\, {g^\prime}^2\,l \nonumber \\
c_{\tilde{t},\tau} & = -\frac{2}{3}\,{g^\prime}^2\,\tau &
a_{\tilde{b},v_1} & = \frac{1}{12}(12\,\lambda_b^2-{g^\prime}^2-3\,g_2^2)\,v_1 &
a_{\tilde{b},v_2} & = \frac{1}{12}({g^\prime}^2+3\,g_2^2)\, v_2  
\nonumber \\
a_{\tilde{b},l} & = - \frac{1}{12}({g^\prime}^2-3\,g_2^2)\,l &
a_{\tilde{b},\tau} & = \frac{1}{6}\,{g^\prime}^2\,\tau &
b_{\tilde{b},v_1} & = \frac{\lambda_b}{\sqrt{2}}\,A_b \nonumber \\
b_{\tilde{b},v_2} & = -\frac{\lambda_b}{\sqrt{2}}\,\mu &
b_{\tilde{b},l} & = \frac{1}{2}\,\lambda_b\,\lambda_\tau\,\tau &
b_{\tilde{b},\tau} & = \frac{1}{2}\,\lambda_b\,\lambda_\tau\, l\nonumber \\
c_{\tilde{b},v_1} & = \frac{1}{6}(6\,\lambda_b^2\,-\,{g^\prime}^2)\,
v_1 & 
c_{\tilde{b},v_2} & = \frac{1}{6}\,{g^\prime}^2\,v_2  &
c_{\tilde{b},l} & = -\frac{1}{6}\,{g^\prime}^2\,l \nonumber \\
c_{\tilde{b},\tau} & = \frac{1}{3}\,{g^\prime}^2\,\tau & 
a_{G^0,l} & = 2\,g_2^2\,\sin^2\,\theta_W\,l & 
a_{G^0,\tau} & = 2\, {g^\prime}^2\,\tau  \nonumber \\
b_{G^0,l} & = g_2^2\,\tan\,\theta_W\,\cos (2\theta_W)\,l &
c_{G^0,v_1} & = \frac{g_2^2}{2\cos^2\theta_W}\,v_1  &
c_{G^0,v_2} & = \frac{g_2^2}{2\cos^2\theta_W}\,v_2 \nonumber \\
c_{G^0,l} & = \frac{g_2^2}{2\cos^2\theta_W}\,\cos (2\theta_W)\,l & & & &
\end{align}
For the charged higgses, we have
\begin{align}
a_{\pm,v_1} & = \frac{1}{4}({g^\prime}^2+g_2^2)\,v_1 & 
a_{\pm,v_2} & = -\frac{1}{4}({g^\prime}^2-g_2^2)\,v_2 &        
a_{\pm,l} & = \frac{1}{4}({g^\prime}^2+g_2^2)\,l \nonumber \\
a_{\pm,\tau} & = \frac{1}{2}(2\,\lambda_\tau^2\,+\,{g^\prime}^2)\,\tau &
b_{\pm,v_1} & = \frac{1}{4}\,g_2^2\,v_2 & 
b_{\pm,v_2} & = \frac{1}{4}\,g_2^2\,v_1 \nonumber \\
c_{\pm,v_1} & = -\frac{1}{4}({g^\prime}^2-g_2^2)\,v_1  &
c_{\pm,v_2} & = \frac{1}{4}({g^\prime}^2+g_2^2)\,v_2 &     
c_{\pm,l} & = -\frac{1}{4}({g^\prime}^2+g_2^2)\,l \nonumber \\
c_{\pm,\tau} & = \frac{1}{2}\,{g^\prime}^2\,\tau & 
d_{\pm,v_1} & = -\frac{1}{4}(2\,\lambda_\tau^2\,-\,g_2^2)\,l & 
d_{\pm,l} & = -\frac{1}{4}(2\,\lambda_\tau^2\,-\,g_2^2)\,v_1
\nonumber  \\
d_{\pm,\tau} & = -\frac{\lambda_\tau}{\sqrt{2}}\,A_\tau &
e_{\pm,v_2} & = \frac{g_2^2}{4}\,l & 
e_{\pm,l} & =  \frac{g_2^2}{4}\,v_2 \nonumber \\
e_{\pm,\tau} & = -\frac{\lambda_\tau}{\sqrt{2}}\,\mu &
f_{\pm,v_1} & = \frac{1}{4}({g^\prime}^2+g_2^2)\,v_1 & 
f_{\pm,v_2} & = -\frac{1}{4}({g^\prime}^2+g_2^2)\,v_2 \nonumber \\ 
f_{\pm,l} & = \frac{1}{4}({g^\prime}^2+g_2^2)\,l &
f_{\pm,\tau} & = \frac{1}{2}(2\,\lambda_\tau^2\,-\,{g^\prime}^2)\,
\tau \;\;\; . & & 
\end{align}
For the pseudoscalars,
\begin{align}
a_{\bar{H},v_1} & = \frac{1}{4}({g^\prime}^2+g_2^2)\,v_1 &
a_{\bar{H},v_2} & = -\frac{1}{4}({g^\prime}^2+g_2^2)\,v_2 &
a_{\bar{H},l} & = \frac{1}{4}(4\,\lambda_\tau^2\,+\,{g^\prime}^2\,-\,g_2^2)\,l 
\nonumber \\
a_{\bar{H},\tau} & = \frac{1}{2}(2\,\lambda_\tau^2\,-\,{g^\prime}^2)
\, \tau & 
c_{\bar{H},v_1} & = -\frac{1}{4}({g^\prime}^2+g_2^2)\,v_1 & 
c_{\bar{H},v_2} & = \frac{1}{4}({g^\prime}^2+g_2^2)\,v_2 \nonumber \\
c_{\bar{H},l} & = -\frac{1}{4}({g^\prime}^2-g_2^2)\,l &
c_{\bar{H},\tau} & = \frac{1}{2}\,{g^\prime}^2\,\tau &
d_{\bar{H},\tau} & = -\frac{\lambda_\tau}{\sqrt{2}}\,A_\tau \nonumber \\
e_{\bar{H},l} & = -\frac{\lambda_\tau}{\sqrt{2}}\,A_\tau &
f_{\bar{H},\tau} & = -\frac{\lambda_\tau}{\sqrt{2}}\,\mu &
g_{\bar{H},l} & = -\frac{\lambda_\tau}{\sqrt{2}}\,\mu \nonumber \\
h_{\bar{H},v_1} & = \frac{1}{4}(4\,\lambda_\tau^2\,+\,{g^\prime}^2\,-\,
g_2^2)\,v_1 & 
h_{\bar{H},v_2} & = -\frac{1}{4}({g^\prime}^2-g_2^2)\,v_2 & 
h_{\bar{H},l} & = \frac{1}{4}({g^\prime}^2+g_2^2)\,l \nonumber \\
h_{\bar{H},\tau} & = \frac{1}{2}(2\,\lambda_\tau^2\,-\,{g^\prime}^2)
\, \tau & 
i_{\bar{H},v_1} & = -\frac{\lambda_\tau}{\sqrt{2}}\,A_\tau & 
i_{\bar{H},v_2} & = \frac{\lambda_\tau}{\sqrt{2}}\,\mu \nonumber \\
j_{\bar{H},v_1} & = \frac{1}{2}(2\,\lambda_\tau^2\,-\,{g^\prime}^2)
\, v_1 & 
j_{\bar{H},v_2} & = \frac{1}{2}\,{g^\prime}^2\,v_2 & 
j_{\bar{H},l} & = \frac{1}{2}(2\,\lambda_\tau^2\,-\,{g^\prime}^2)\,
l \nonumber \\
j_{\bar{H},\tau} & = {g^\prime}^2\,\tau \;\;\; .  & & & & 
\end{align}
And for the higgs scalars,
\begin{align}
a_{H,v_1} & = \frac{3}{4}({g^\prime}^2+g_2^2)\,v_1 &
a_{H,v_2} & = -\frac{1}{4}({g^\prime}^2+g_2^2)\,v_2 &
a_{H,l} & = \frac{1}{4}(4\,\lambda_\tau^2\,+\,{g^\prime}^2\,-\,g_2^2)\,l 
\nonumber \\
a_{H,\tau} & = \frac{1}{2}(2\,\lambda_\tau^2\,-\,{g^\prime}^2)\,\tau &
b_{H,v_1} & = -\frac{1}{4}({g^\prime}^2+g_2^2)\,v_2 & 
b_{H,v_2} & = -\frac{1}{4}({g^\prime}^2+g_2^2)\,v_1 \nonumber \\
c_{H,v_1} & = -\frac{1}{4}({g^\prime}^2+g_2^2)\,v_1 & 
c_{H,v_2} & = \frac{3}{4}({g^\prime}^2+g_2^2)\,v_2 &
c_{H,l} & = -\frac{1}{4}({g^\prime}^2-g_2^2)\,l \nonumber \\
c_{H,\tau} & = \frac{1}{2}\,{g^\prime}^2\,\tau &
d_{H,v_1} & = \frac{1}{4}(4\,\lambda_\tau^2\,+\,{g^\prime}^2-g_2^2)\,l & 
d_{H,l} & = \frac{1}{4}(4\,\lambda_\tau^2\,+\,{g^\prime}^2-g_2^2)\,v_1
\nonumber \\
d_{H,\tau} & = \frac{\lambda_\tau}{\sqrt{2}}\,A_\tau &
e_{H,v_1} & = \frac{1}{2}(2\,\lambda_\tau^2\,-\,{g^\prime}^2)\,\tau &
e_{H,l} & = \frac{\lambda_\tau}{\sqrt{2}}\,A_\tau \nonumber \\
e_{H,\tau} & = \frac{1}{2}(2\,\lambda_\tau^2\,-\,{g^\prime}^2)\,v_1 &
f_{H,v_2} & = -\frac{1}{4}\,({g^\prime}^2\,-\,g_2^2)\,l & 
f_{H,l} & = -\frac{1}{4}\,({g^\prime}^2\,-\,g_2^2)\,v_2 \nonumber \\
f_{H,\tau} & = -\frac{\lambda_\tau}{\sqrt{2}}\,\mu & 
g_{H,v_2} & = \frac{1}{2}\,{g^\prime}^2\,\tau & 
g_{H,l} & = - \frac{\lambda_\tau}{\sqrt{2}}\,\mu \nonumber \\ 
g_{H,\tau} & = \frac{1}{2}\,{g^\prime}^2\,v_2 & 
h_{H,v_1} & = \frac{1}{4}(4\,\lambda_\tau^2\,+\,{g^\prime}^2\,-\,g_2^2)\,v_1 & 
h_{H,v_2} & = -\frac{1}{4}({g^\prime}^2-g_2^2) \,v_2 \nonumber \\ 
h_{H,l} & = \frac{3}{4}({g^\prime}^2+g_2^2)\,l & 
h_{H,\tau} & = \frac{1}{2}(2\,\lambda_\tau^2\,-\,{g^\prime}^2)\,\tau &
i_{H,v_1} & = \frac{\lambda_\tau}{\sqrt{2}}\,A_\tau \nonumber \\ 
i_{H,v_2} & = -\frac{\lambda_\tau}{\sqrt{2}}\,\mu & 
i_{H,l} & = \frac{1}{2}(2\,\lambda_\tau^2\,-\,{g^\prime}^2)\,\tau &
i_{H,\tau} & = \frac{1}{2}(2\,\lambda_\tau^2\,-\,{g^\prime}^2)\, l \nonumber \\
j_{H,v_1} & = \frac{1}{2}(2\,\lambda_\tau^2\,-\,{g^\prime}^2)\,v_1 & 
j_{H,v_2} & = \frac{1}{2}\,{g^\prime}^2\,v_2 & 
j_{H,l} & = \frac{1}{2}(2\,\lambda_\tau^2\,-\,{g^\prime}^2)\,l 
\nonumber \\
j_{H,\tau} & = 3\,{g^\prime}^2\,\tau \;\;\; .  & & & & 
\end{align}
For the charginos, from eq.~\eqref{eq:char}, we have
\begin{eqnarray}
A_{\chi^\pm,(v_1,v_2,l)} & = & g_2^2\,(v_1\;,\;v_2\;,\;l) \hspace{2cm}
A_{\chi^\pm,\tau} \; = \; \lambda_\tau^2\,\tau \nonumber \\
B_{\chi^\pm,v_1} & = & \frac{1}{2}\,g_2^4\,(v_2^2+l^2)\,v_1\;-\;\frac{
\lambda_\tau}{\sqrt{2}}\,g_2^2\,M_2\,l\,\tau\;-\;g_2^2\,\mu\,M_2\,v_2 \nonumber 
\\
B_{\chi^\pm,v_2} & = & \frac{1}{2}\,g_2^4\,v_1^2\,v_2\;+\;\frac{\lambda_\tau^2}{
2}\,g_2^2\,\tau^2\,v_2\;-\;\frac{\lambda_\tau}{\sqrt{2}}\,g_2^2\,\mu\,l\,\tau
\;-\;g_2^2\,\mu\,M_2\,v_1\nonumber \\
B_{\chi^\pm,l} & = & g_2^2\,\left(\mu^2\,+\,\frac{g_2^2}{2}\,v_1^2\right)\,l
\;-\;\frac{\lambda_\tau}{\sqrt{2}}\,g_2^2\,(\mu\,v_2\,+\,M_2\,v_1)\,\tau 
\nonumber \\
B_{\chi^\pm,\tau} & = & \lambda_\tau^2\,\left(M_2^2\,+\,\frac{g_2^2}{2}\,v_2^2
\right)\,\tau\;-\;\frac{\lambda_\tau}{\sqrt{2}}\,g_2^2\,(\mu\,v_2\,+\,M_2\,v_1)
\,l \;\;\; .
\end{eqnarray}
Finally, for the neutralinos, from eq.~\eqref{eq:neut}, we find
\begin{eqnarray}
B_{\chi^0,v_1} & = & \frac{1}{2}\,(2\,\lambda_\tau^2\,+\,
{g^\prime}^2\,+\,g_2^2)\,v_1 \hspace{1.5cm} B_{\chi^0,v_2} \; = \; 
\frac{1}{2}\,({g^\prime}^2\,+\,g_2^2)\,v_2 \nonumber \\
B_{\chi^0,l} & = & \frac{1}{2}\,(2\,\lambda_\tau^2\,+\,
{g^\prime}^2\,+\,g_2^2)\,l\hspace{1.7cm} B_{\chi^0,\tau} \; = \; (\lambda_\tau^2
\,+\,2\,{g^\prime}^2)\,\tau \nonumber \\
C_{\chi^0,v_1} & = & -\,\left[2\,\lambda_\tau^2\,(M_1\,+\,M_2)\;+\;(
{g^\prime}^2\,M_2\,+\,g_2^2\,M_1)\right]\,\frac{v_1}{2}\;+\;({g^\prime}^2\,+\,
g_2^2)\,\frac{\mu}{2}\,v_2 \;- \nonumber \\
 & & \frac{\lambda_\tau}{2\sqrt{2}}\,(3\,{g^\prime}^2\,-\,g_2^2\,-\,2\,
\lambda_\tau^2)\,l\,\tau \nonumber \\
C_{\chi^0,v_2} & = & -\,({g^\prime}^2\,M_2\,+\,g_2^2\,M_1)\,\frac{v_2}{2}\;+\;
({g^\prime}^2\,+\,g_2^2)\,\frac{\mu}{2}\,v_1 \nonumber \\
C_{\chi^0,l} & = & -\,\left[2\,\lambda_\tau^2\,(M_1\,+\,M_2)\;+\;(
{g^\prime}^2\,M_2\,+\,g_2^2\,M_1)\right]\,\frac{l}{2}\;- \nonumber \\
 & & \frac{\lambda_\tau}{2\sqrt{2}}\,(3\,{g^\prime}^2\,-\,g_2^2\,-\,2\,
\lambda_\tau^2)\,v_1\,\tau \nonumber \\
C_{\chi^0,\tau} & = & -\,\left[\lambda_\tau^2\,(M_1\,+\,M_2)\;+\;2\,{g^\prime}^2
\,M_2\right]\,\tau\;-\;\frac{\lambda_\tau}{2\sqrt{2}}\,(3\,{g^\prime}^2\,-\,
g_2^2\,-\,2\,\lambda_\tau^2)\,v_1\,l \nonumber \\
D_{\chi^0,v_1} & = & \frac{\lambda_\tau^3}{\sqrt{2}}\,(M_1\,+\,M_2)\,l\,\tau\;+
\;\lambda_\tau^2\,\left[\frac{}{}\right. {g^\prime}^2\,\tau^2\,+\,\frac{1}{4}\,
({g^\prime}^2\,+\, g_2^2)\,(2\,v_1^2\,+\,v_2^2\,-\,2\,l^2)\,- \nonumber \\
 & & \left. \frac{}{} M_1\,M_2\,+\,\mu^2\right]\,v_1\;+\;\frac{\lambda_\tau}{2
\sqrt{2}}\,(g_2^2\,M_1\,-\,3\,{g^\prime}^2\,M_2)l\,\tau\;+ \nonumber \\   
 & & \frac{1}{2}\,\left[{g^\prime}^2\,g_2^2\,\tau^2\,v_1\;+\;({g^\prime}^2\,M_2
\,+\,g_2^2\,M_1)\,\mu\,v_2\right] \nonumber \\
D_{\chi^0,v_2} & = & \frac{\lambda_\tau^2}{4}\,({g^\prime}^2\,+\,g_2^2)\,(v_1^2
\,+\,l^2\,+\,\tau^2)\,v_2\;+\;\frac{\lambda_\tau}{2\sqrt{2}}\,({g^\prime}^2\,-\,
g_2^2)\,\mu\,l\,\tau\;+\;\frac{1}{2}\,\left[\frac{}{}\right. {g^\prime}^2\,g_2^2
\,\tau^2\,v_2\;+ \nonumber \\
 & & \left. ({g^\prime}^2\,M_2\,+\,g_2^2\,M_1)\,\mu\,v_1\right] \nonumber \\
D_{\chi^0,l} & = & \frac{\lambda_\tau^3}{\sqrt{2}}\,(M_1\,+\,M_2)\,v_1\,\tau\;+
\;\lambda_\tau^2\,\left[{g^\prime}^2\,\tau^2\,+\,\frac{1}{4}\,({g^\prime}^2\,+\,
g_2^2)\,(2\,l^2\,+\,v_2^2\,-\,2\,v_1^2)\,- \right. \nonumber \\
 & & \left. \frac{}{} M_1\,M_2\right]\,l\;+\;\frac{\lambda_\tau}{2\sqrt{2}}\,
\left[(g_2^2\,M_1\,-\,3\,{g^\prime}^2\,M_2)\,v_1 \,+\,({g^\prime}^2\,-\,g_2^2)\,
\mu\,v_2\right]\,\tau\;+ \nonumber \\
 & & \frac{l}{2}\,\left[{g^\prime}^2\,g_2^2\,\tau^2\,l \,+\,({g^\prime}^2\,+\,
g_2^2)\,\mu^2\right] \nonumber \\
D_{\chi^0,\tau} & = & \frac{\lambda_\tau^3}{\sqrt{2}}\,(M_1+M_2)\,v_1\,l\,+
\, \lambda_\tau^2\,\left[{g^\prime}^2\,(v_1^2+l^2+2\,\tau^2)+
\frac{1}{4}\,({g^\prime}^2\,+\,g_2^2)\,v_2^2 - M_1\,M_2\right]\,\tau\,+
\nonumber \\
 & & \frac{\lambda_\tau}{2\sqrt{2}}\,\left[(g_2^2\,M_1\,-\,3\,{g^\prime}^2\,M_2)
\,v_1\,+\,({g^\prime}^2\,-\,g_2^2)\,\mu\,v_2\right]\,l\;+ \nonumber \\
 & & \frac{\tau}{2}\,\left[{g^\prime}^2\,g_2^2\,(v_1^2\,+\,v_2^2\,+\,l^2)\,+\,4
\,{g^\prime}^2\,\mu^2\right] \nonumber \\
E_{\chi^0,v_1} & = & \frac{\lambda_\tau^3}{4\sqrt{2}}\,\left[({g^\prime}^2\,+\,
g_2^2)\,v_2^2\;-\;4\,M_1\,M_2\right]\,l\,\tau\;+\;\frac{\lambda_\tau^2}{4}\,
\left[ ({g^\prime}^2\,M_2\,+\,g_2^2\,M_1)\,(2\,l^2\,- \right. \nonumber \\
 & & \,2\,v_1^2\,-\,v_2^2)\,v_1 \,-\,({g^\prime}^2\,+\,g_2^2)\,\mu\,(l^2\,-\,
3\,v_1^2)\,v_2\;-\;4\,(M_1\,+\,M_2)\,\mu^2\,v_1\;- \nonumber \\
 & & \left. 4\,{g^\prime}^2\,\left(M_2\,
v_1\,-\,\frac{\mu}{2}\,v_2\right) \right]\;+\;\frac{\lambda_\tau}{2\sqrt{2}}\,
\left({g^\prime}^2\,g_2^2\,\tau^2\;-
\;2\,{g^\prime}^2\,\mu^2\right)\,l\,\tau\;+\;{g^\prime}^2\,g_2^2\,\frac{\mu}{2}
\,v_2\,\tau^2 \nonumber \\
E_{\chi^0,v_2} & = & \frac{\lambda_\tau^3}{2\sqrt{2}}\,({g^\prime}^2\,+\,g_2^2)
\,v_1\,v_2\,l\,\tau\;+\;\frac{\lambda_\tau^2}{4}\,\left[-({g^\prime}^2\,M_2\,+\,
g_2^2\,M_1)\,(v_1^2\,+\,l^2\,+\,\tau^2)\,v_2\;- \right. \nonumber \\
 & & \left. ({g^\prime}^2\,+\,g_2^2)\,\mu\,(l^2\,-v_1^2)\,v_1\;+\;2\,
{g^\prime}^2\,\mu\,v_1\,\tau^2\right]\;+\;\frac{\lambda_\tau}{2\sqrt{2}}\,(g_2^2
\,M_1\,-\,{g^\prime}^2)\,\mu\,l\,\tau \;+ \nonumber \\
 & &  {g^\prime}^2\,g_2^2\,\frac{\mu}{2}\,v_1\,\tau^2 \nonumber \\
E_{\chi^0,l} & = & \frac{\lambda_\tau^3}{4\sqrt{2}}\,\left[({g^\prime}^2\,+\,
g_2^2)\,v_2^2\;-\;4\,M_1\,M_2\right]\,v_1\,\tau\;+\;\frac{\lambda_\tau^2}{4}\,
\left[-({g^\prime}^2\,M_2\,+\,g_2^2\,M_1)\,(2\,l^2\,-\right. \nonumber \\
 & & 2 \left. \,v_1^2\,+\,v_2^2)\,l\;-\;4\,{g^\prime}^2\,M_2\,\tau^2\;-\;2\,(
{g^\prime}^2\,+\,g_2^2)\,\mu\,v_1\,v_2\,l
\right]\;+ \nonumber \\
 & & \frac{\lambda_\tau}{2\sqrt{2}}\,\left[{g^\prime}^2\,(g_2^2\,\tau^2\,
-\,\mu^2)\,v_1\;+\;\mu\,(g_2^2\,M_1\,-\,{g^\prime}^2\,M_2)\,v_2\right]\,\tau\;-
\; \frac{\mu^2}{2}\,({g^\prime}^2\,M_2\,+\,g_2^2\,M_1)\,l \nonumber \\
E_{\chi^0,\tau} & = & \frac{\lambda_\tau^3}{4\sqrt{2}}\,\left[({g^\prime}^2\,+\,
g_2^2)\,v_2^2\;-\;4\,M_1\,M_2\right]\,v_1\,l\;+\;\frac{\lambda_\tau^2}{4}\,
\left[-({g^\prime}^2\,M_2\,+\,g_2^2\,M_1)\,v_2^2\,\tau\;- \right. \nonumber \\
 & & 4\,{g^\prime}^2\,M_2 \,(v_1^2\,+\,l^2\,+\,2\,\tau^2)\;+\;\left. 4\,
{g^\prime}^2\,\mu\,v_1\,v_2\tau\right]\;+\;\frac{\lambda_\tau}{2\sqrt{2}}\,
\left[(3\,{g^\prime}^2\,g_2^2\,\tau^2\,-\,2\,{g^\prime}^2\,\mu^2)\,v_1\;+ 
\right. \nonumber \\
 & & \left. (g_2^2\,M_1\,-\,{g^\prime}^2\,M_2)\,\mu\,v_2 \right]\,l\;+\;
{g^\prime}^2\,(g_2^2\,v_1\,v_2\,-\,2\,\mu\,M_2)\,\mu\,\tau \nonumber \\
F_{\chi^0,v_1} & = & -\frac{\lambda_\tau^3}{4\sqrt{2}}\,({g^\prime}^2\,M_2\,+\,
g_2^2\,M_1)\,v_2^2\,l\,\tau\;+\;\frac{\lambda_\tau^2}{4}\,\left[({g^\prime}^2\,
M_2\,+\,g_2^2\,M_1)\,\mu\,(l^2\,-\,3\,v_1^2)\,v_2\;- \right. \nonumber \\
 & & \left. 2\,{g^\prime}^2\,\mu\,M_2
\,v_2\,\tau^2\;+\;4\,\mu^2\,M_1\,M_2\,v_1\right]\;+\;\frac{\lambda_\tau}{
\sqrt{2}}\,{g^\prime}^2\,\mu^2\,M_2\,l\,\tau \nonumber \\
F_{\chi^0,v_2} & = & -\frac{\lambda_\tau^3}{2\sqrt{2}}\,({g^\prime}^2\,M_2\,+
\, g_2^2\,M_1)\,v_1\,v_2\,l\,\tau\;+\;\frac{\lambda_\tau^2}{4}\,\left[
({g^\prime}^2\,M_2\,+\,g_2^2\,M_1)\,\mu\,(l^2\,-\,v_1^2)\,v_1\;- \right. 
\nonumber \\
 & & \left. {g^\prime}^2 \,\left(2\,\mu\,M_2\,v_1\;+\;g_2^2\,v_2\,\tau^2\right)
\,\tau^2\right]\;+\;
\frac{\lambda_\tau}{2\sqrt{2}}\,{g^\prime}^2\,g_2^2\,\mu\,l\,\tau^3\nonumber \\
F_{\chi^0,l} & = & -\frac{\lambda_\tau^3}{4\sqrt{2}}\,({g^\prime}^2\,M_2\,+\, 
g_2^2\,M_1)\,v_1\,v_2^2\,\tau\;+\;\frac{\lambda_\tau^2}{2}\,({g^\prime}^2\,M_2\,
+\,g_2^2\,M_1)\,\mu\,v_1\,v_2\,l\;+ \nonumber \\
 & & \frac{\lambda_\tau}{2\sqrt{2}}\,{g^\prime}^2\,(2\,\mu\, M_2\,v_1\,+\,g_2^2
\,v_2\,\tau^2)\,\mu\,\tau \;-\;\frac{\mu^2}{2}\,{g^\prime}^2\, g_2^2\,l\,\tau^2 
\nonumber \\
F_{\chi^0,\tau} & = & -\frac{\lambda_\tau^3}{4\sqrt{2}}\,({g^\prime}^2\,M_2\,+\,
g_2^2\,M_1)\,v_1\,v_2^2\,l\;-\;\frac{\lambda_\tau^2}{2}\,{g^\prime}^2\,(2\,\mu\,
M_2\,v_1\,+\,g_2^2\,\tau^2\,v_2)\,v_2\,\tau\;+ \nonumber \\
 & & \frac{\lambda_\tau}{2\sqrt{2}}\,{g^\prime}^2\,(2\,\mu\,M_2\,v_1\,+\,3\,
g_2^2\,v_2\,\tau^2)\,\mu\,l\;-\;\frac{\mu^2}{2}\,{g^\prime}^2\,g_2^2\,l^2\,\tau
\end{eqnarray}

\end{document}